\documentclass[prb,floatfix,twocolumn,showpacs,superscriptaddress]{revtex4}
\usepackage{graphicx,times}
\usepackage{amssymb,amsmath}

\def\et{{\it et al.}}
\def\dimer{$(1/\sqrt{2}) (\vert\!\!\uparrow \downarrow \rangle - \vert\!\!\downarrow \uparrow \rangle)$}

\begin{document}

\author{Matthieu Mambrini}

\affiliation{Laboratoire de Physique Th\'eorique, CNRS \& Universit\'e de Toulouse,
  F-31062 Toulouse, France}

\date{\today}
\title{Hardcore dimer aspects of the SU(2) Singlet wavefunction}

\begin{abstract}

We demonstrate that any SU(2) singlet wavefunction can be characterized by a set of Valence Bond occupation numbers, testing dimer presence/vacancy on pairs of sites. This genuine quantum property of singlet states (i) shows that SU(2) singlets share some of the intuitive features of hardcore quantum dimers, (ii) gives rigorous basis for interesting albeit apparently ill-defined quantities introduced recently in the context of Quantum Magnetism or Quantum Information to measure respectively spin correlations and bipartite entanglement and, (iii) suggests a scheme to define
consistently a wide family of quantities analogous to high order spin correlation.  This result is demonstrated in the framework of a general functional mapping between the Hilbert space generated by an arbitrary number of spins and a set of algebraic functions found to be an efficient analytical tool for the description of quantum spins or qubits systems.

\end{abstract}
\maketitle

\section{Introduction}

Coupling a large number of spin $s=1/2$ to a total spin $S=0$ state (singlet) leads to a very rich class of wavefunctions which are the central pieces of many physical problems ranging from Condensed Matter to Quantum Information. Indeed, the singlet state plays a major role in Quantum Magnetism especially in the low temperature properties of SU(2) Heisenberg frustrated antiferromagnets~: below the finite spin gap energy scale, the physics is dominated by rather exotic singlet ground states (valence bond crystal, plaquettes state, spin-liquid, \ldots) and often not so well understood low lying singlet excitations~\cite{Misguich}. It also attracted considerable interest since Anderson suggested that the Resonating Valence Bond (singlet) state (RVB) is the relevant insulating parent state of high-$T_c$ superconductivity~\cite{Anderson}. Interestingly, these genuine quantum objects naturally emerge in the different context of Quantum Information because they are somehow paradigmatic of entanglement. Indeed, the simplest singlet ($S=0$) wavefunction, namely the dimer state \dimer, is a maximally entangled state made out two spin $1/2$. For large systems, singlet wavefunction can naturally lead to massive entanglement~\cite{Bose}.

\begin{figure}
\centerline{\includegraphics*[angle=0,width=0.9\linewidth]{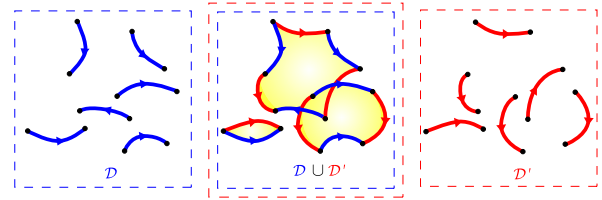}}
  \caption{\label{fig:vb}
Two VB states corresponding to the coverings $\cal D$ and ${\cal D}'$. The overlap
diagram obtained by the superimposition $\cal D\cup {\cal D}'$ involves a set a closed loops.}
\end{figure}

Obtaining a simple and tractable description of singlet states when the number of spins becomes large, is thus a crucial issue.
In that respect, it has been shown since the early days of quantum mechanics that Valence Bond (VB) states~\cite{Rumer} are a much more convenient
and elegant approach to this question than expansions of singlets as linear combinations of $S_i^z$ eigenstates. Indeed,  Hulth\'en showed~\cite{Hulthen} that any singlet state can be written as a (non unique) linear combination of VB states, i.e. arbitrary range coverings of the $N$ sites with dimers~:
\begin{equation}
\label{eq:rvb}
\vert \varphi_{\cal D} \rangle = \bigotimes_{(i,j) \in {\cal D}} [i,j],
\end{equation}
where $[i,j] = (1/\sqrt{2}) (\vert\!\!\uparrow_i \downarrow_j \rangle - \vert\!\!\downarrow_i \uparrow_j \rangle) $ and ${\cal D}$ is a dimer covering of the system, defined as a
partition of the $\{ 1,\ldots,N\}$ ensemble into $N/2$ {\it oriented}
couples of sites $\{(i_1,j_1),\ldots,(i_{N/2},j_{N/2})\}$ (see figure \ref{fig:vb}).

However, while emphasizing the role of an apparently intuitive source of spin correlation and entanglement (namely the ``bond dimer'' \dimer), this approach leads to a non-orthogonal and massively overcomplete~\cite{Sizes} description of singlet states that, in turn, destroys any interpretation of dimers as hardcore objects~\cite{Fuzziness}. Paradoxically enough, this conceptually problematic ``dimer fuzziness'' (caused by the intrinsic non-orthogonality of these objects) does not lead to major inconsistencies when neglected. Let us select three significant examples, from 
the contexts of Quantum Magnetism and Quantum Information, where this apparently ill-defined notion of bond occupation appears either implicitly or explicitly and is shown  to give yet interesting and consistent results~: (i) The Liang-Dou\c{c}ot-Anderson (LDA) wavefunction~\cite{LDA} convincingly reinforces the idea that spin correlations can be driven by individual VB strength tuning, and in the same spirit Sandvik~\cite{SandvikMC} obtained very consistent results for the VB length distribution while mentioning the potential definition problem, (ii) The Rokhsar-Kivelson quantum dimer model~\cite{Hardcore} that considers dimers as hardcore and orthogonal objects is widely accepted as a good starting point to the study of SU(2) highly frustrated (or more generally gapped) Heisenberg antiferromagnets~\cite{HardcoreExample1,HardcoreExample2}, (iii) Alet \et~\cite{AletEntropy} recently introduced a VB Entanglement Entropy numerically shown to capture all the features of the von Neumann Entanglement Entropy but defined as if dimers where hardcore and orthogonal objects.

In this article, we use a general functional framework to describe any singlet wavefunctions that allows to demonstrate
that a VB occupation number testing the presence/absence of a SU(2) dimer on a given pair of spins can be defined and computed unambiguously as an intrinsic property of any SU(2) singlet state. The implications are threefold~: (i) We show that SU(2) singlets actually share some of the intuitive hardcore-like features of quantum dimers, (ii) We give rigorous basis for relevant albeit apparently ill-defined quantities introduced recently in the context of Quantum Magnetism such as VB Length Distribution~\cite{SandvikMC} or Quantum Information such as VB Entanglement Entropy~\cite{AletEntropy} and give a systematic scheme to define consistently a wide family of quantities analogous to high order spin correlation, (iii) The functional mapping provides a convenient framework for further studies of correlated and entangled quantum spin or qubits systems and can be extended to arbitrary values of spin $s$ and  total spin $S$.

The paper is organized as follow: in the first section we introduce and demonstrate the validity of the functional mapping. In the second part the VB occupation number is explicitly defined and shown to be the parent quantity for VB Length Distribution~\cite{SandvikMC} and VB Entanglement Entropy~\cite{AletEntropy}. The final section is devoted to the generalization of the VB occupation number concept to arbitrary order VB correlation functions and the extension to higher spin $s$ and total spin $S$ values of the functional mapping.

\section{Functional mapping}

We introduce on each site $i$ of the $N$-spin system a real (or complex) variable $x_i$ and associate to any dimer covering $\cal D$ the $N$-variable function $\varphi_{\cal D}$ defined by~:
\begin{equation}
\label{eq:rep}
\varphi_{\cal D}(x_1,\ldots,x_N) = \prod_{(i,j) \in {\cal D}} d(x_i,x_j),
\end{equation}
where $d(x_i,x_j)$ is a function of $x_i$ and $x_j$  that will be determined later. Each function $\varphi_{\cal D}(\{x_i\})$ of the mapping corresponds to a pure VB state $\vert \varphi_{\cal D} \rangle$ defined by (\ref{eq:rvb}). The key properties of the VB states are their non-orthogonality and overcompleteness, both being closely related.  Let us us recall the overlap rule that allows to compute the overlap~\cite{Sutherland} between two VB configurations ${\cal D}$ and ${\cal D}'$,
\begin{equation}
\label{eq:overlap}
{\cal O}_{{\cal D},{{\cal D}'}}=\langle \varphi_{\cal D} \vert \varphi_{{\cal D}'} \rangle = \varepsilon_{{\cal D} , {{\cal D}'}} . 2^{n_l-N/2},
\end{equation}
$n_l$ being the number of closed loops in the overlap diagram obtained by superimposing both configurations ${\cal D}$ and ${\cal D}'$ and $\varepsilon_{{\cal D} , {{\cal D}'}}$ the sign coming from the relative orientations of dimers in ${\cal D}$ and ${\cal D}'$ (see figure \ref{fig:vb}). The overcompleteness of the basis implies its non-orthogonality and on the other hand, all the linear relations
between VB states are implicitly encoded into the overlap matrix ${\cal O}$~: Because any singlet state can be expressed as a linear combination of arbitrary range VB states, the rank of ${\cal O}$ is the singlet subspace size ${\cal N}_{0} = N! / ((1+N/2)! (N/2)!)$. Its size being ${\cal N} = N! / (2^{N/2} (N/2)!)$, all the overcompleteness of the basis is described by the ${\cal N} - {\cal N}_{0}$ independent singular eigenvectors of ${\cal O}$. Hence, in order to obtain a faithful functional representation, it is necessary and sufficient to determine
$d(x_i,x_j)$ and define a scalar product $\langle\langle \varphi_{\cal D}(\{x_i\}) \vert\vert \varphi_{{\cal D}'}(\{x_i\}) \rangle\rangle$ acting on these functions  that mimics the usual scalar product for VB states~: $\langle\langle \varphi_{\cal D}(\{x_i\}) \vert\vert \varphi_{{\cal D}'}(\{x_i\}) \rangle\rangle =
{\cal O}_{{\cal D},{{\cal D}'}}$.

Let us show that $d(x_i,x_j)$ can be chosen as
\begin{equation}
\label{eq:dimer_rep}
d(x_i,x_j) = \frac{1}{\sqrt{2}} (x_i-x_j),
\end{equation}
and
\begin{equation}
\label{eq:scalar_product_rep}
\langle\langle \varphi_{\cal D}(\{x_i\}) \vert\vert \varphi_{{\cal D}'}(\{x_i\}) \rangle\rangle =
\sum_{x_i= \pm 1/2} \varphi_{\cal D}(\{x_i\}) \times \varphi_{{\cal D}'}(\{x_i\}).
\end{equation}

Given the form (\ref{eq:rep}), $\varphi_{\cal D}(\{x_i\}) \times \varphi_{{\cal D}'}(\{x_i\})$ involves bond terms
and each variable $x_i$ appears exactly twice. Hence, this term can be represented graphically by the 
set of even size closed loops of the overlap diagram (see figure \ref{fig:vb}). Loops being disconnected, 
$\langle\langle \varphi_{\cal D}(\{x_i\}) \vert\vert \varphi_{{\cal D}'}(\{x_i\}) \rangle\rangle$ is just a product
of the individual loop ($l$) traces contributions, 
\begin{equation}
\label{eq:trace_contrib}
\langle\langle \varphi_{\cal D}(\{x_i\}) \vert\vert \varphi_{{\cal D}'}(\{x_i\}) \rangle\rangle = \prod_{l \in {\cal D} \cup {\cal D}'} t_l
\end{equation}
with, for a loop of size $p$,
\begin{equation}
\label{eq:loop_contrib}
t_l = \sum_{x_i = \pm 1/2} d(x_{i_1},x_{i_2}) d(x_{i_2},x_{i_3}) \ldots d(x_{i_p},x_{i_1}).
\end{equation}
Choosing (\ref{eq:dimer_rep}) implies that only two sequences, $\{x_i\}=\{\pm 1/2,\mp 1/2,\ldots,\mp 1/2\}$, give a non vanishing contribution to $t_l$. As a consequence, $t_l=2.2^{-p(l)/2}.\varepsilon_l$ with $p(l)$ the size (number of sites) of the loop and $\varepsilon_l = \pm 1$ the contribution of the loop $l$ to the global sign $\varepsilon_{{\cal D} , {{\cal D}'}} = \prod_{l \in {\cal D} \cup {\cal D}'} \varepsilon_l$. 
Because the total size of the loops is just the number of sites $N$, the overlap $\langle\langle \varphi_{\cal D}(\{x_i\}) \vert\vert \varphi_{{\cal D}'}(\{x_i\}) \rangle\rangle = \varepsilon_{{\cal D} , {{\cal D}'}} 2^{n_l-N/2}$ as expected.

Using (\ref{eq:scalar_product_rep}), we showed that the set of functions
\begin{equation}
\label{eq:rep2}
\varphi_{\cal D}(x_1,\ldots,x_N) = \frac{1}{2^{N/4}} \prod_{(i,j) \in {\cal D}} (x_i-x_j),
\end{equation}
have the same linear properties as the VB states (\ref{eq:rvb}) and thus provides an algebraic mapping
of the singlet subspace~\cite{MapExample}.

We emphasize that this mapping is general as it establishes a one to one correspondence between any VB state (\ref{eq:rvb}) and its functional counterpart (\ref{eq:rep2})~: any subclass of VB states maps to its corresponding subclass both defining isomorphic Hilbert spaces.
In the next section, we show (i) the particular role of the subset made of bipartite VB states, (ii) that valence bond occupation number is unambiguously defined using bipartite VB states despite non-orthogonality and overcompleteness.

\begin{figure}
\centerline{\includegraphics*[angle=0,width=0.9\linewidth]{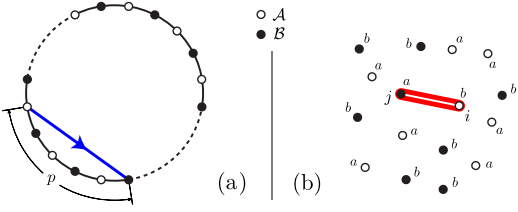}}
  \caption{\label{fig:bipartite}
(a) Arrangement of $N$ spins around a circle, alternating ${\cal A}$ sites and ${\cal B}$ sites. (b) In the definition (\ref{eq:count_operator}) of bond $(i,j)$ occupation operator, the limit for $x_k$ is taken to $a=1/2$ for ${\cal A}$ sites and $b \neq a$ for $\cal B$ sites except for $x_i$ and $x_j$ for which the inverse convention  is used~: $a$ if $i$ (or $j$) is part of ${\cal B}$ and $b$ if $i$ (or $j$) is part of ${\cal A}$.}
\end{figure} 


\section{Valence Bond occupation numbers}

{\it Bipartition.} The overcompleteness of the full VB basis prevents to define consistently a VB occupation number. We now show that such a definition is possible by restraining expansions of singlet states to a subset of $\cal V$, yet overcomplete and spanning all the  singlet sector.
Let us partition the $N$-spin system into two $N/2$-spin subsets denoted $\cal A$ and $\cal B$~\cite{Bipartition}. 
We consider the  subset ${\cal V}^{\text{b}}$ of ${\cal V}$ containing the ${\cal N}^{\text{b}} = (N/2)!$ bipartite VB states~:
\begin{equation}
\label{eq:bip_rvb}
\vert \varphi_{\cal D} \rangle = \bigotimes_{\substack{(i,j) \in {\cal D} \\ i \in {\cal A}, j \in {\cal B}}} [i,j].
\end{equation}
${\cal V}^{\text{b}}$ also provides an overcomplete description of the singlet subspace~: let us arrange the $N$ spins on a circle
in such a way that ${\cal A}$ sites alternate with ${\cal B}$ sites (see figure \ref{fig:bipartite} (a)) and choose an arbitrary singlet state $\vert \psi \rangle$. As shown by Rumer \et~\cite{Rumer}, there exist at least one decomposition of $\vert \psi \rangle$ as a linear combination of {\em non-crossing}~\cite{Noncrossing} VB states. A simple inspection at figure \ref{fig:bipartite} (a) shows that these non-crossing states are also bipartite. The occurence of a non-bipartite dimer would lead to the existence of an odd sized arc, hence causing at least one crossing. As a consequence any singlet state $\vert \psi \rangle$ can be written as,
\begin{equation}
\label{eq:bip_decomposition}
\vert \psi \rangle = \sum_{{\cal D} \in {\cal V}^{\text{b}}} \lambda_{\cal D} \vert \varphi_{\cal D} \rangle.
\end{equation}
This decomposition is not unique due to ${\cal N}^{\text{b}} - {\cal N}$ independent linear relations that reflect the overcompleteness of ${\cal V}^{\text{b}}$.

{\it Bond Occupation.} We denote by $n_{(i,j)}$ the occupation number of a bond $(i,j)$ (bipartite or not) in a bipartite VB state $\vert \varphi_{\cal D} \rangle$ (\ref{eq:bip_rvb}) and define it as,
\begin{equation}
\label{eq:bond_occupation_pure}
n_{(i,j)}(\vert \varphi_{\cal D} \rangle)=
	\begin{cases}
	1& \text{if $(i,j)$ belongs to $\cal D$},\\
	0& \text{otherwise}.
\end{cases}
\end{equation}
For any singlet state $\vert \psi \rangle$ decomposed on ${\cal V}^{\text{b}}$ as (\ref{eq:bip_decomposition}), we show that the bond occupation
\begin{equation}
\label{eq:bond_occupation_mixing}
n_{(i,j)}(\vert \psi \rangle)= \sum_{{\cal D} \in {\cal V}^{\text{b}}} \lambda_{\cal D} \; n_{(i,j)}(\vert \varphi_{\cal D} \rangle),
\end{equation}
is {\em independent} of the decomposition (\ref{eq:bip_decomposition}) used, hence providing a consistent intrinsic definition of a bond occupation number
for any singlet state. This statement is valid if any relation of overcompleteness implies its bond occupation counterpart~:
\begin{equation}
\label{eq:overcompleteness}
\sum_{{\cal D} \in {\cal V}^{\text{b}}} \lambda_{\cal D} \; \vert \varphi_{\cal D} \rangle = 0 \;\; \Rightarrow \;\; \sum_{{\cal D} \in {\cal V}^{\text{b}}} \lambda_{\cal D} \; n_{(i,j)}(\vert \varphi_{\cal D} \rangle) = 0.
\end{equation}

To demonstrate this point, let us use the functional mapping presented in the previous section and introduce the operator ${\cal D}_{(i,j)}$ acting on the functions of the mapping,
\begin{equation}
\label{eq:count_operator}
{\cal D}_{(i,j)} f(\{x_k\}) = 2^{N/4} \lim_{\substack{x_{i} \rightarrow \alpha, x_{j} \rightarrow \beta \\ x_{k\in {\cal A}} \rightarrow 1/2 \\ x_{k\in {\cal B}} \rightarrow -1/2}} (\partial_{x_i}-\partial_{x_j}) f(\{x_k\}),
\end{equation}
with $\alpha$ or $\beta=-1/2$ (resp. $+1/2$) if $i$ or $j\in{\cal A}$ (resp. $\cal B$) (see figure \ref{fig:bipartite} (b)).
One can easily check that ${\cal D}_{(i,j)}$ have the following properties~: (i) Linearity~: ${\cal D}_{(i,j)} (f+g) = {\cal D}_{(i,j)} f+{\cal D}_{(i,j)} g$, (ii) For any ${\cal D} \in {\cal V}^{\text{b}}$, ${\cal D}_{(i,j)} \varphi_{\cal D} = 1$ if $(i,j) \in {\cal D}$ and ${\cal D}_{(i,j)} \varphi_{\cal D} = 0$ if $(i,j) \notin {\cal D}$. Using (\ref{eq:rep2}), l.h.s. of equation (\ref{eq:overcompleteness}) becomes,
\begin{equation}
\label{eq:overcompleteness_mapping}
\sum_{{\cal D} \in {\cal V}^{\text{b}}} \lambda_{\cal D} \; \prod_{\substack{(p,q) \in {\cal D} \\ p \in {\cal A}, q \in {\cal B}}} (x_p-x_q) = 0,
\end{equation}
and applying ${\cal D}_{(i,j)}$ to (\ref{eq:overcompleteness_mapping}) immediately leads to the r.h.s. of (\ref{eq:overcompleteness}) which concludes the demonstration.

At this point, it is important to mention that $n_{(i,j)}$ is reminiscent of a bond correlation $C_{ij} = (-4/3){\bf S}_i . {\bf S}_j$ but is \emph{essentially different}. Indeed, $n^2_{(i,j)}(\vert \psi_{\cal D} \rangle)=\langle \varphi_{\cal D} \vert C_{ij} \vert \varphi_{\cal D} \rangle$ for any pure VB state. But, in the generic case of a superposition (\ref{eq:bip_decomposition}), both notions decouple due to the off-diagonal terms induced by the non-orthogonality of VB states~:
\begin{equation}
\label{eq:not_correlation}
\langle \psi \vert C_{ij} \vert \psi \rangle = n^2_{(i,j)}(\vert \psi \rangle) + \sum_{\substack{{\cal D},{\cal D}' \\ {\cal D} \neq {\cal D}'}} \lambda_{\cal D} \lambda_{{\cal D}'}  \langle \varphi_{{\cal D}'} \vert C_{ij} \vert \varphi_{\cal D} \rangle.
\end{equation}

{\it Illustration.} For $N=6$ (then ${\cal N}_0 = 5$,  ${\cal N} = 15$ and ${\cal N}^{\text{b}} = 6$) we define the bipartite VB states taking ${\cal A} = \{1,3,5\}$ and ${\cal B} = \{2,4,6\}$. The overcomplete bipartite VB basis is~: $\vert \alpha \rangle=[1,2][3,4][5,6]$, $\vert \beta \rangle=[1,6][3,2][5,4]$, $\vert \gamma \rangle=[1,4][5,2][3,6]$, $\vert \delta \rangle=[1,2][3,6][5,4]$, $\vert \varepsilon \rangle=[1,4][3,2][5,6]$ and $\vert \zeta \rangle=[1,6][5,2][3,4]$. The sole relation of overcompleteness writes $\vert \alpha \rangle$+$\vert \beta \rangle$+$\vert \gamma\rangle$-$\vert \delta \rangle$-$\vert \varepsilon \rangle$-$\vert \zeta\rangle=0$. We can check that e.g. the occupation of bond $(5,2)$ on the singlet state  $3\vert \alpha \rangle + 2\vert \zeta \rangle$ that can be directly evaluated to $3 \times 0 + 2 \times 1 = 2$ is indeed independent of the decomposition~: $3\vert \alpha \rangle + 2\vert \zeta \rangle = 5\vert \alpha \rangle +2\vert \beta \rangle+2\vert \gamma\rangle - 2\vert \delta \rangle -2 \vert \varepsilon \rangle$ leads to $5\times 0 + 2\times 0 + 2\times 1 - 2\times 0 - 2\times 0 =2$. On the other hand, the expectation value of $(-4/3) {\bf S}_2 . {\bf S}_5$ is 5 and not $4=2^2$ for this state.

{\it VB Length Distribution \& VB Entanglement Entropy.}  Using this result any quantity built out from bond occupation/vacancy scheme on a singlet state $\vert \psi \rangle$ can be defined unambiguously. The VB Length Distribution computed in \cite{SandvikMC} writes,
\begin{equation}
\label{eq:VB_length1}
{\cal P}_l (\vert \psi \rangle) = \frac{{\cal L}_l (\vert \psi \rangle)}{\sum_{l} {\cal L}_l (\vert \psi \rangle)} \;\; {\text{with}} \;\;
{\cal L}_l (\vert \psi \rangle) = \sum_{\substack{(i,j) {\text{ with}} \\ {\text{length }} l}} {\cal D}_{(i,j)}  (\vert \psi \rangle),
\end{equation}
where ${\cal L}_l (\vert \psi \rangle)$ is the total occupation of bonds with length $l$. A definition of the VB Entanglement Entropy introduced in~\cite{AletEntropy}
as an equivalent of von Neumann Entanglement Entropy measurement of the bipartite entanglement between subsystems $\Omega$ and $\bar{\Omega}$ spanning the all system is,
\begin{equation}
\label{eq:VB_EE}
{\cal S}^{\text{VB}}_{\Omega,\bar{\Omega}}(\vert \psi \rangle) = \ln 2 \times  \sum_{\substack{(i,j) {\text{ crossing}}  \\ \Omega \vert \bar{\Omega} }} {\cal D}_{(i,j)}  (\vert \psi \rangle).
\end{equation}

\section{Extensions and Generalizations}

{\it VB correlation functions.} Comparing (\ref{eq:dimer_rep}) and (\ref{eq:count_operator}) exhibit a duality in the mapping between the dimer representation $(x_i-x_j)$ and a dimer susceptibility $(\partial_{x_i}-\partial_{x_j})$. It is then straightforward to introduce $p-$order VB occupation numbers (or {\em VB correlation functions}), similar to $2p-$spin correlation functions, by defining {\em linear} operators in the spirit of (\ref{eq:count_operator}) containing more than one bond susceptibility $(\partial_{x_{i_1}}-\partial_{x_{j_1}})\ldots(\partial_{x_{i_p}}-\partial_{x_{j_p}})$. The analogous of equation (\ref{eq:not_correlation}) for such an operator, shows that both quantities are equivalent for a pure VB state but becomes distinct yet consistently defined for a VB superposition. While much simpler to compute, we infer that VB correlation functions provide the same insights than usual correlations functions when computed on a given singlet state.

{\it Higher total spin sectors for $s=1/2$.} Following Hulth\'en's scheme~\cite{Hulthen} higher total spin $S$ sectors with fixed total $S_z$ can be (overcompletely) represented by introducing in the coverings (\ref{eq:rvb})
an arbitrary range totally symmetrized state of $2S$ spins $\{i_1,\ldots,i_{2S}\}_{S_z}$. The functional mapping can be extended to these case in the spirit of (\ref{eq:dimer_rep}) using the generating function ${\cal G}^S (\mu,\{x_{i_p}\}) = \prod_{p=1}^{2S} \left ( 1+\mu x_{i_p} \right )$ and  representing $\{i_1,\ldots,i_{2S}\}_{S_z}$ with
\begin{equation}
\label{eq:symmetric}
d^{S}_{S_z}(x_{i_1},\ldots, x_{i_{2S}})=  \sqrt{ \frac{2^{2 S_z}}{(2S)!} \frac{(S-S_z)!}{(S+S_z)!} } \left ( \partial_\mu^{(S_z+S)} {\cal G}^S \right ) \Big |_{\mu=0}.
\end{equation}
In particular $d^{1}_{0}(x_1,x_2)=(1/\sqrt{2})(x_i+x_j)$ represents the $S_z=0$ bond triplet. Mixed total spin state such as the N\'eel state can be represented as well in this framework~: If we denote ${\cal A}$ the $N/2$-site sublattice with up spins, it is easy to check that the corresponding N\'eel state is represented by $\Pi_{\cal A}=\prod_{i \in {\cal A}} x_i$ since, using (\ref{eq:scalar_product_rep}), $\langle\langle \varphi_{\cal D}(\{x_i\}) \vert\vert \Pi_{\cal A} \rangle\rangle =(1/2)^{N/4}$ for any bipartite $\cal D$.

{\it Singlet sector for higher spins $s$.} As noticed by Tasaki~\cite{Tasaki}, VB concept can be extended to higher values of $s$~: $[i,j]_s = \frac{1}{\sqrt{2s+1}} \sum_{s_z=-s}^{+s} (-1)^{s-s_z} \vert -s_z,+s_z\rangle$.
The overlap ${\cal O}_{{\cal D},{{\cal D}'}}$ (\ref{eq:overlap}) becomes $\langle \varphi_{\cal D} \vert \varphi_{{\cal D}'} \rangle = \varepsilon_{{\cal D} , {{\cal D}'}} . (2s+1)^{n_l-N/2}$ with $\varepsilon_{{\cal D} , {{\cal D}'}}=1$ for integer $s$. 
It can be readily checked that choosing $d_s (x_i,x_j) = \frac{1}{\sqrt{2s+1}} (-1)^{s-x_i} \delta_{x_i,-x_j}$
and enlarging the support of the sum defining the scalar product (\ref{eq:scalar_product_rep}) to $\sum_{x_i=-s}^{x_i=+s}$
leads to $\langle\langle \varphi_{\cal D}(\{x_i\}) \vert\vert \varphi_{{\cal D}'}(\{x_i\}) \rangle\rangle = \langle \varphi_{\cal D} \vert \varphi_{{\cal D}'} \rangle$ thus providing the extension of the mapping. The proof that $s$-spins VB states provide an overcomplete description of the singlet sector (if $N$ is large enough to ensure that the number of coverings is larger than size of the singlet sector) is beyond the scope of this article~\cite{Mambrini_prep}.

\section{Conclusions}

Any SU(2) singlet state can be characterized by a set of bond occupation numbers~: despite the fuzziness induced
by the overcompleteness of the VB basis, singlet wavefunctions conserve intuitive hardcore dimer features such as the possibility
of deciding whether a SU(2) dimer live on a bond or not. This genuine quantum property formalizes the suggestive idea behind the LDA wavefunction~\cite{LDA} that VB bond strength can drive the correlations properties of the singlet wavefunction and gives insights to the question of the interplay between non-orthogonality, overcompleteness and hardcore properties. Moreover, this allows consistent definitions of quantities relevant for Quantum Magnetism (VB Length Distribution~\cite{SandvikMC} and VB correlations~\cite{Beach3D} of any order) or Quantum Information (VB Entanglement Entropy~\cite{AletEntropy}) emphasizing the intuitive role of dimers as elementary bricks to build correlations or entanglement. We note that all these quantities are particularly suitable for numerical calculation in the framework introduced by Sandvik~\cite{SandvikMC}. From a methodological point of view, the functional mapping used to derive this result appears as a suggestive tool to describe quantum spin or qubits systems and can be extended naturally to higher spins $s$ and higher total spin $S$ sectors.

I thank Fabien Alet and Sylvain Capponi for very interesting discussions. This work was supported by the Agence Nationale
de la Recherche (France).

\vfill

\begin{thebibliography}{}
\bibitem{Sizes} Indeed, the number of states in $\cal V$ is ${\cal N} = N! / (2^{N/2} (N/2)!) \sim (N/e)^{N/2}$ which is much larger than the singlet subspace size ${\cal N}_{0} = N! / ((1+N/2)! (N/2)!) \sim 2^N / \sqrt{N}$.
\bibitem{Fuzziness} Bond $(1,2)$ is occupied in the 4-spin singlet state $[1,2][3,4]$. But $[1,2][3,4]=[1,3][2,4]+[1,4][3,2]$, and the state can be expressed as a linear combination of VB states, none of them having bond $(1,2)$ occupied.
\bibitem{MapExample} As an illustration, one can easily check that the simplest overcompleteness relation
$[1,2][3,4]+[1,3][4,2]+[1,4][2,3]=0$ is mapped the algebraic identity $(x_1-x_2)(x_3-x_4)+(x_1-x_3)(x_4-x_2)+(x_1-x_4)(x_2-x_3)=0$.
\bibitem{Bipartition} Such a choice is
always possible, since we do not presume any underlying geometry, lattice structure nor hamiltonian for the system.
\bibitem{Noncrossing} A VB state is non-crossing for a given arrangement of sites around a circle if its representation contains no dimer intersection.
\end{thebibliography}
\end{document}